\documentclass[1p]{elsarticle}
\usepackage{bm,amsmath,amsfonts,amssymb}
\usepackage[english,french]{babel}
\newcommand{\ud}{\mathrm{d}}
\newcommand{\xt}{\bm{x},t}
\newcommand{\xvt}{\bm{x},\bm{v},t}
\newcommand{\xyt}{\bm{x},\bm{y},t}
\newcommand{\inttx}{\!\int_0^{\mathcal{T}}\!\frac{\ud t}{\mathcal{T}}\!\int_{\mathbb{P}}\!\ud\bm{x}\,}
\newcommand{\inttxv}{\!\int_0^{\mathcal{T}}\!\frac{\ud t}{\mathcal{T}}\!\int_{\mathbb{P}}\!\ud\bm{x}\!\int_{\mathbb{R}^d}\!\ud\bm{v}\,}
\newcommand{\itxv}{\!\int_{t,\bm{x},\bm{v}}}

\begin{document}

 \selectlanguage{english}
 
 \begin{frontmatter}

  \title{Settling~velocity of~quasi-neutrally-buoyant~inertial~particles}

  \author{Marco Martins Afonso}
  \ead{marcomartinsafonso@hotmail.it (Corresponding author, telephone +351 2204022 59, facsimile +351 2204022 09)}
  \author{S\'{\i}lvio M.\ A.\ Gama}
  \address{Centro de Matem\'atica da Universidade do Porto, Rua do Campo Alegre 687, 4169-007 Porto, Portugal}

  \begin{abstract}
   We investigate the sedimentation properties of quasi-neutrally buoyant inertial particles carried by incompressible zero-mean fluid flows.
   We obtain generic formulae for the terminal velocity in generic space-and-time periodic (or steady) flows,
   along with further information for flows endowed with some degree of spatial symmetry such as odd parity in the vertical direction. 
   These expressions consist in space-time integrals of auxiliary quantities which satisfy partial differential equations
   of the advection--diffusion--reaction type, that can be solved at least numerically since our scheme
   implies a huge reduction of the problem dimensionality from the full phase space to the classical physical space.
  \vskip 0.5\baselineskip
%  \selectlanguage{francais}
  \noindent{\bf R\'esum\'e}
  \vskip 0.5\baselineskip
  \noindent
   Nous \'etudions les propri\'et\'es de s\'edimentation de particules inertielles dot\'ees de flottabilt\'e quasi neutre et transport\'ees par un
   \'ecoulement incompressible \`a moyenne nulle. Nous obtenons des formules g\'en\'eriques pour la vitesse terminale dans des \'ecoulements en g\'en\'eral
   p\'eriodiques en espace et en temps (ou statiques), avec d'ult\'erieures informations disponibles pour les \'ecoulements dot\'ees de sym\'etries
   spatiales sp\'ecifiques telles que parit\'e n\'egative dans la direction verticale. Ces expressions consistent en int\'egraux spatio--temporels
   de quantit\'es auxiliaires qui ob\'eissent \`a des \'equations aux d\'eriv\'ees partielles du type advection--diffusion--r\'eaction. Ces derni\`eres
   peuvent \^etre r\'esolues au moins num\'eriquement car notre proc\'edure implique une forte r\'eduction de la dimensionalit\'e du probl\`eme,
   de l'espace des phases complet \`a l'espace physique classique.
  \vskip 0.5\baselineskip
%  \selectlanguage{english}
  \end{abstract}

 \end{frontmatter}

 \section{Introduction}
 
 Particles advected by a fluid are called ``inertial'' if, when studying their motion, one cannot neglect the particle relative inertia with respect to
 the surrounding fluid. This is usually due to their (small but) not negligible size, and/or to a mismatch between the two mass densities.
 Common examples are represented by small bubbles in liquids, droplets in gases, and aerosols in a generic fluid.
 The comprehension of the dynamics of these impurities is still an open issue from the theoretical, experimental and numerical points of view
 \cite{R90,R92,BFF01,B03,WM03,FP04,ML04,CBBBCLMT06,VCVLMPT08}. Implications are relevant in many applied domains: plankton dynamics in biology
 \cite{KPSTT00}; chemical reactors, spray combustion and emulsions in industrial engineering \cite{HDS08}; planet formation in astrophysics \cite{MM02};
 transport of pollutants or floaters, rain initiation and sedimentation processes in geophysics \cite{FFS02}.
 
 Our focus is precisely on sedimentation, with a special attention to those situations where the mass-density ratio is (different from but) very close to
 unity. This is for instance the case for most living beings suspended in an aquatic medium. The intuitive picture is the following: inertia causes a
 deviation of the particles from the underlying fluid trajectory, which leads to inhomogeneities for the particle concentration in regions of the flow with
 different dynamical properties, due to the presence of symmetry-breaking forces and preferential directions --- in our case, gravity along the vertical.
 Moreover, we will also consider the effect of Brownian diffusivity. This latter is usually neglected in most investigations on inertial particles,
 assuming that Brownian noise is very small for finite-size particles. However, this is not true for tiny particles, and especially in biophysical
 applications, where a limited capacity of autonomous movement could be kept into account in this simple fashion. This work therefore represents
 a complementary study with respects to similar ones which focused on the limits of small inertia or of large Brownian diffusivity.

 Our principal objective is to obtain an Eulerian description of the settling (i.e., falling or rising) in steady or periodic flows starting from the
 well-known Lagrangian viewpoint for particle motion.
 Despite this, our theory provides the whole detailed statistical information of particle motion. Indeed, the probability density
 function of having a particle in a given position at a certain time is available from our approach, at least in a perturbative way.
 However, this implies the resolution of partial differential equations, which in general can be accomplished only numerically.

 The paper is organized as follows. In section \ref{sec:eq}
 we define the problem under investigation, we specify our assumptions and we sketch our analytical procedure.
 We enounce the final result for generic flows in section \ref{sec:gf},
 and we specialize it for vertically-antisymmetric ones in section \ref{sec:pf}.
 Conclusions and perspectives follow in section \ref{sec:cp}. The %appendix
 \ref{sec:ap} is devoted to showing the details of the calculation and to recalling the mathematical tools employed. 
 
 \section{Equations} \label{sec:eq}
 
 We consider a very dilute suspension of point-like inertial particles subject to the gravitational acceleration $\bm{g}$ and to Brownian diffusion,
 carried by a fluid flow.
 We suppose that our $d$-dimensional incompressible velocity field is steady or periodic in time (with period $\mathcal{T}$),
 and periodic in space with unit cell $\mathbb{P}$ of linear size $\ell$.
 It is not a restriction to focus on velocity fields whose average vanishes over $\mathbb{P}$:
 \begin{equation} \label{u}
  \!\int_{\mathbb{P}}\!\ud\bm{x}\,\bm{u}(\xt)=\bm{0}\;.
 \end{equation}
 In this way, any deviation of the settling velocity with respect to the value found in still fluids
 will represent a genuine interplay between gravity and the other properties of particle and flow,
 and not a mere streaming or sweeping effect.
 The same technique can be extended to handle the case of a random, homogeneous, and stationary velocity field \cite{MY75}
 with some non-trivial modifications in the rigorous proofs of convergence \cite{AM91}.
 For an interesting investigation of the role played by mean currents on the eddy diffusivity of tracers,
 see e.g.\ \cite{HM94,MV97,MMV05,CMMV06,FN10,MAMG16}.
 
 Neglecting any possible interaction with other particles or with physical boundaries, and taking into account the
 feedback on the transporting fluid in an effective way by means of a simplified added-mass effect, the Lagrangian
 dynamics reduces to the following set of stochastic differential equations for the particle
 position $\bm{\mathcal{X}}(t)$ and covelocity $\bm{\mathcal{V}}(t)$ \cite{MR83,G83}:
 \begin{equation} \label{basic}
  \left\{\begin{array}{rcl}
   \dot{\bm{\mathcal{X}}}(t)\!\!&\!\!=\!\!&\!\!\bm{\mathcal{V}}(t)+\beta\bm{u}(\bm{\mathcal{X}}(t),t)+\sqrt{2\mathcal{D}}\bm{\mu}(t)\;,\\
   \dot{\bm{\mathcal{V}}}(t)\!\!&\!\!=\!\!&\!\!\displaystyle-\frac{\bm{\mathcal{V}}(t)-(1-\beta)\bm{u}(\bm{\mathcal{X}}(t),t)}{\tau}+(1-\beta)\bm{g}+\frac{\sqrt{2\kappa}}{\tau}\bm{\nu}(t)\;.
  \end{array}\right.
 \end{equation}
 The independent vectorial white noises $\bm{\mu}(t)$ and $\bm{\nu}(t)$ influence the particle dynamics by means of the coupling constants $\mathcal{D}$
 and $\kappa$, which can be identified as Brownian diffusivities \cite{R88}. The presence of two different parameters in the equations for the position
 and the velocity will become clear shortly.
 The pure number $\beta\equiv3\rho_{\mathrm{f}}/(\rho_{\mathrm{f}}+2\rho_{\mathrm{p}})\in[0,3]$, built from the constant
 fluid ($\rho_{\mathrm{f}}$) and particle ($\rho_{\mathrm{p}}$) mass densities, is dubbed ``added-mass factor'' because it takes into account the fact that
 any particle motion necessarily implies some fluid motion around it, thus increasing the intrinsic inertia --- with the sole exception of
 very heavy particles such as aerosols or droplets in a gas ($\beta\simeq0$). It also induces a macroscopic discrepancy between the particle velocity
 $\dot{\bm{\mathcal{X}}}(t)$ and covelocity $\bm{\mathcal{V}}(t)$, which is maximum for very light particles such as bubbles in a liquid ($\beta\simeq3$).
 Alternatively, in terms of slip velocity --- defined as the difference between the particle velocity and the local instantaneous fluid velocity sampled
 by the particle: $\bm{\mathcal{Y}}(t)\equiv\dot{\bm{\mathcal{X}}}(t)-\bm{u}(\bm{\mathcal{X}}(t),t)$ --- the covelocity turns out to be
 $\bm{\mathcal{V}}(t)=\bm{\mathcal{Y}}(t)+(1-\beta)\bm{u}(\bm{\mathcal{X}}(t),t)-\sqrt{2\mathcal{D}}\bm{\mu}(t)$.
 Finally, the Stokes time $\tau$ in the drag term expresses the typical response delay of particles to flow variations, and is defined as
 $\tau\equiv Q^2/(3\gamma\beta)$ for spherical inertial particles of radius $Q$ immersed in a fluid with kinematic viscosity $\gamma$.
 Note however that, as customary in inertial-particle studies, $\beta$ and $\tau$ are assumed as independent parameters, since the latter can be varied
 even when the former is kept fixed by suitably changing $Q$ and $\gamma$.
 The dynamical system (\ref{basic}) neglects the classical contributions due to Basset (time integration for memory/history/wake effects), Oseen
 (nonlinear finite-Reynolds-number correction to the basic Stokes flow), Fax\'en (spatial expansion of the fluid flow for finite particle size)
 and Saffman (lateral lift in case of rotation).

 After statistical averaging of (\ref{basic}) on $\bm{\mu}(t)$ and $\bm{\nu}(t)$ \cite{C43,G85,R89,V07}, the generalized Fokker--Planck (or Kramers, or
 forward Kolmogorov) equation for the phase-space density $p(\xvt)$ is obtained:
 \begin{equation} \label{gfpkfk}
  \left\{\partial_t+\bm{\partial}_{\bm{x}}\cdot[\bm{v}+\beta\bm{u}(\xt)]+\bm{\partial}_{\bm{v}}\cdot\left[\frac{(1-\beta)\bm{u}(\xt)-\bm{v}}{\tau}+(1-\beta)\bm{g}\right]-\mathcal{D}\partial^2_{\bm{x}}-\frac{\kappa}{\tau^2}\partial^2_{\bm{v}}\right\}p=0\;.
 \end{equation}
 Let us denote by $\mathcal{L}(\xvt)$ the linear operator in curly braces on the left-hand side of (\ref{gfpkfk}), so that $\mathcal{L}p=0$.
 For future use, let us also introduce the corresponding physical-space concentration, obtained by integrating on the covelocity variable:
 \begin{equation} \label{pP}
  q(\xt)\equiv\!\int_{\mathbb{R}^d}\!\ud\bm{v}\,p(\xvt)\;.
 \end{equation}

 The particle terminal velocity \cite{MC86,M87a,M87b,WM93,FK02,RMP04,MA08} is defined as a weighted average of the particle velocity, from the first
 equation in (\ref{basic}):
 \begin{equation} \label{dabliu}
  \bm{w}\equiv\langle\bm{\mathcal{V}}(t)+\beta\bm{u}(\bm{\mathcal{X}}(t),t)+\sqrt{2\mathcal{D}}\bm{\mu}(t)\rangle_{p}=\inttxv[\bm{v}+\beta\bm{u}(\xt)]p(\xvt)
 \end{equation}
 (here and in what follows, the average on the temporal period $\mathcal{T}$ is skipped for steady flows). Notice that in general such quantity corresponds
 to a mean behavior and not to an asymptotic value --- except for the case of still fluids if Brownian diffusion is negligible.
 Indeed, inside a flow each particle can wander in any direction and follow more or less closely the underlying fluid trajectory,
 but the overall evolution of a bunch of non-interacting particles will consist an a falling/rising described by $\bm{w}$.
 On the contrary, in our model, the well-known ``bare'' asymptotic value of sedimentation in still fluids is:
 \begin{equation} \label{ast}
  \bm{\mathcal{W}}\equiv(1-\beta)\tau\bm{g}\;.
 \end{equation}
 As proven in %section
 \ref{sec:pr}, the deviation of the terminal velocity from its bare value can be rewritten using (\ref{pP}) as:
 \begin{equation} \label{delta}
  \bm{\mathcal{Z}}\equiv\bm{w}-\bm{\mathcal{W}}=\inttxv\bm{u}(\xt)p(\xvt)=\inttx\bm{u}(\xt)q(\xt)\;.
 \end{equation}
 
 Now, let us focus on particles whose mass density differs only slightly (either in excess or in shortfall) from the fluid one \cite{BCPP00,MFS07,SH08}.
 Since $\beta\simeq1$, then $1-\beta$ is small but with an undefined sign, so we introduce a second small parameter in the form of
 $\alpha\equiv|1-\beta|\ll1$. We also define $\mathcal{J}\equiv\mathrm{sgn}(1-\beta)$, thus $\beta=1-\mathcal{J}\alpha$. It can be shown that in this
 situation it is possible to proceed analytically only if one makes the further assumption that the Brownian-diffusion coefficient $\kappa$ appearing in
 the equation for the particle acceleration be small as well, namely with the same asymptotic behavior as the mass-density mismatch:
 $\kappa\sim\alpha\ll1$; or, in other words, one can define a finite constant $\mathcal{K}\equiv\kappa/|1-\beta|=\alpha^{-1}\kappa$ with dimensions of
 square length over time. Notice that no assumption is made on the Brownian diffusivity $\mathcal{D}$ driving the particle velocity, which can then be
 thought of as a regularizing parameter. As is well known, the diffusivity of a tracer particle --- obeying to (\ref{basic}) with $\tau=0$ --- would turn
 out to be simply $\mathcal{D}+\kappa$, but for inertial particles the situation is more subtle and, indeed, our analytical procedure works only if
 $\kappa$ is small and $\mathcal{D}$ is non-zero. It is also worth mentioning that, had one proceeded on a Lagrangian route before turning to the
 (Eulerian) phase-space description, the zeroth-order situation $\beta=1$ would correspond to a Markovian process driven by a colored noise
 (Ornstein--Uhlenbeck) in the Langevin equation, as already described in \cite{CC99}. The Lagrangian approach has also been followed in \cite{BMM17} to
 find exact expressions for the particle eddy diffusivity in shear or Gaussian flows.
 
 Upon rescaling the covelocity variable according to $\bm{v}\mapsto\bm{y}\equiv\bm{v}/\sqrt{|1-\beta|}=\alpha^{-1/2}\bm{v}$, the generalized
 Fokker--Planck operator splits into:
 \begin{equation*}
  \mathcal{L}=\mathcal{L}^{(0)}+\alpha^{1/2}\mathcal{L}^{(1)}+\alpha\mathcal{L}^{(2)}\;,
 \end{equation*}
 with
 \begin{subequations} \label{l012}
  \begin{eqnarray}
   \mathcal{L}^{(0)}\!\!&\!\!=\!\!&\!\!\partial_t+\bm{u}(\xt)\cdot\bm{\partial}_{\bm{x}}-\mathcal{D}\partial^2_{\bm{x}}-\tau^{-1}\bm{\partial}_{\bm{y}}\cdot\bm{y}-\mathcal{K}\tau^{-2}\partial^2_{\bm{y}} \label{l012a}\\
   \mathcal{L}^{(1)}\!\!&\!\!=\!\!&\!\!\bm{y}\cdot\bm{\partial}_{\bm{x}}+\mathcal{J}[\tau^{-1}\bm{u}(\xt)+\bm{g}]\cdot\bm{\partial}_{\bm{y}} \label{l012b}\\
   \mathcal{L}^{(2)}\!\!&\!\!=\!\!&\!\!-\mathcal{J}\bm{u}(\xt)\cdot\bm{\partial}_{\bm{x}}\;. \label{l012c}
  \end{eqnarray}
 \end{subequations}
 For the sake of notational simplicity, we define a ``gravitational velocity field'' $\bm{z}(\xt)\equiv\bm{u}(\xt)+\tau\bm{g}$ (with
 $\bm{\partial}_{\bm{x}}\cdot\bm{u}=0\Rightarrow\bm{\partial}_{\bm{x}}\cdot\bm{z}=0$) and two linear operators,
 \begin{equation} \label{m}
  \mathcal{M}(\xt)\equiv\partial_t+\bm{u}(\xt)\cdot\bm{\partial}_{\bm{x}}-\mathcal{D}\partial^2_{\bm{x}}
 \end{equation}
 (advection--diffusion in physical space) and
 \begin{equation} \label{n}
  \mathcal{N}(\bm{y})\equiv\bm{\partial}_{\bm{y}}\cdot\bm{y}+\mathcal{K}\tau^{-1}\partial^2_{\bm{y}}
 \end{equation}
 (related to the Ornstein--Uhlenbeck formalism). In terms of them,
 \begin{equation*}
   \mathcal{L}^{(0)}=\mathcal{M}(\xt)-\tau^{-1}\mathcal{N}(\bm{y})\;,\qquad\mathcal{L}^{(1)}=\bm{y}\cdot\bm{\partial}_{\bm{x}}+\mathcal{J}\tau^{-1}\bm{z}(\xt)\cdot\bm{\partial}_{\bm{y}}\;.
 \end{equation*}
 Our rescaling is dictated by the close analogy with the situation described in \cite{MA08,MAM11,MAMM12}, where the small-inertia limit was performed.
 In that case the small quantity at denominator was the square root of $\tau$, while here it is that of $|1-\beta|$.
 As shown in the appendix, the advantage of such a rescaling lies in the fact that it allows for a full decoupling of the rescaled covelocity from the
 physical-space dynamics, and for the resolution of equations based on the operator (\ref{n}) in terms of a basic Gaussian state.
 Note that in the present framework we have to request the smallness of $\kappa$ explicitly, a condition which on the contrary was somehow implicit
 in those works, as explained in \cite{MAM11} by introducing the non-dimensional Stokes and P\'eclet numbers (whose product was required to be $O(1)$).
 
 It is now natural to expand the phase-space density into a power series in $\sqrt{\alpha}$ and to replace into (\ref{gfpkfk}):
 \begin{equation*}
  p(\xyt)=\sum_{\mathcal{I}=0}^{\infty}\alpha^{\mathcal{I}/2}p^{(\mathcal{I})}(\xyt)\;,
 \end{equation*}
 implying that
 \begin{subequations} \label{01K}
  \begin{eqnarray}
   \mathcal{L}^{(0)}p^{(0)}\!\!&\!\!=\!\!&\!\!0 \label{01Ka}\\
   \mathcal{L}^{(0)}p^{(1)}\!\!&\!\!=\!\!&\!\!-\mathcal{L}^{(1)}p^{(0)} \label{01Kb}\\
   \mathcal{L}^{(0)}p^{(\mathcal{I})}\!\!&\!\!=\!\!&\!\!-\mathcal{L}^{(1)}p^{(\mathcal{I}-1)}-\mathcal{L}^{(2)}p^{(\mathcal{I}-2)}\qquad(\mathcal{I}\ge2)\;. \label{01Kc}
  \end{eqnarray}
 \end{subequations}

 \section{Results for periodic incompressible flows} \label{sec:gf}

 The terminal velocity is accordingly expanded as:
 \begin{equation} \label{wK}
  \bm{w}=\sum_{\mathcal{I}=0}^{\infty}\alpha^{\mathcal{I}/2}\bm{w}^{(\mathcal{I})}\;,\qquad\bm{\mathcal{Z}}=\sum_{\mathcal{I}=0}^{\infty}\alpha^{\mathcal{I}/2}\bm{\mathcal{Z}}^{(\mathcal{I})}\;.
 \end{equation}
 Since $\bm{\mathcal{W}}=\alpha\mathcal{J}\tau\bm{g}$, then
 $\bm{w}^{(\mathcal{I})}=\bm{\mathcal{Z}}^{(\mathcal{I})}+\delta_{\mathcal{I}2}\mathcal{J}\tau\bm{g}$.
 It can be shown (see appendix for details) that actually all the half-integer orders of these expressions (corresponding to odd $\mathcal{I}$) identically
 vanish, so that in practice such expansions reduce to common analytical ones. Moreover, one also sees that $\bm{w}^{(0)}=\bm{0}=\bm{\mathcal{Z}}^{(0)}$,
 i.e.\ particles with exactly-neutral buoyancy --- which would macroscopically stand still in fluids at rest --- on average do not settle either in the
 presence of our class of flows. In what follows, we are going to provide the expressions for the terminal velocity up to the second order,
 that is $\bm{w}^{(2)}$ and $\bm{w}^{(4)}$. Formula (\ref{dabliu}) can be
 manipulated in order to succeed in performing the covelocity integrals, and what is left are space--time integrals of a set of fields satisfying
 equations of the advection--diffusion--reaction type in the configuration space. At working order, such fields of our interest are denoted by
 $q^{(0)}$, $r^{(1)}_i$, $q^{(2)}$, $s^{(2)}_{ij}$, $r^{(3)}_i$ and $q^{(4)}$.
 Apart from imposing the constancy of $q^{(0)}=\ell^{-d}$, their other partial differential equations are solvable analytically
 only for specific flows such as parallel ones. However, such a class of flow is not relevant for our scope, since no contribution to the terminal
 velocity arises from them. Nevertheless, our procedure allows for at least a numerical resolution in generic flows, because of the
 drastic reduction in the dimensionality of the problem from $2d+1$ to $d+1$.
 
 Postponing all details to %section
 \ref{sec:ap}, and defining $\nabla_i\equiv\partial_{x_i}$, we assert first of all that:
 \begin{equation} \label{w2}
  w^{(2)}_i=\mathcal{J}\tau g_i+\mathcal{Z}^{(2)}_i\;,\qquad\mathcal{Z}^{(2)}_i=\inttx u_i(\xt)q^{(2)}(\xt)\;,
 \end{equation}
 with $q^{(2)}$ introduced in (\ref{o2}); and
 \begin{equation} \label{w4}
  w^{(4)}_i=\mathcal{Z}^{(4)}_i=\inttx u_i(\xt)q^{(4)}(\xt)\;,
 \end{equation}
 with $q^{(4)}$ introduced in (\ref{o4}).\\
 To determine the order $\alpha^1$, the set of relevant equations consists of:
 \begin{subequations} \label{ale}
  \begin{eqnarray}
   (\mathcal{M}+\tau^{-1})r^{(1)}_i\!\!&\!\!=\!\!&\!\!\ell^{-d}\mathcal{J}\mathcal{K}^{-1}z_i \label{aleA}\\{}
   \mathcal{M}q^{(2)}\!\!&\!\!=\!\!&\!\!-\mathcal{K}\tau^{-1}\nabla_i r^{(1)}_i \label{aleB}\;,
  \end{eqnarray}
 \end{subequations}
 with $r^{(1)}_i$ introduced in (\ref{o1}).\\
 To analyze $O(\alpha^2)$ too, the system also comprises:
 \begin{subequations} \label{set}
  \begin{eqnarray}
   (\mathcal{M}+2\tau^{-1})s^{(2)}_{ij}\!\!&\!\!=\!\!&\!\!-\nabla_ir^{(1)}_j+\mathcal{J}\mathcal{K}^{-1}z_ir^{(1)}_j \label{setA}\\{}
   (\mathcal{M}+\tau^{-1})r^{(3)}_i\!\!&\!\!=\!\!&\!\!-\nabla_iq^{(2)}+\mathcal{J}\mathcal{K}^{-1}z_iq^{(2)}+\mathcal{J}\bm{u}\cdot\bm{\nabla}r^{(1)}_i-\mathcal{K}\tau^{-1}\nabla_j(s^{(2)}_{ji}+s^{(2)}_{ij})\ \label{setB}\\{}
   \mathcal{M}q^{(4)}\!\!&\!\!=\!\!&\!\!\mathcal{J}\bm{u}\cdot\bm{\nabla}q^{(2)}-\mathcal{K}\tau^{-1}\nabla_ir^{(3)}_i \label{setC}\;.
  \end{eqnarray}
 \end{subequations}
 with $s^{(2)}_{ij}$ and $r^{(3)}_i$ introduced in (\ref{o2}) and (\ref{o3}) respectively.\\
 The conclusions that can be drawn analytically at this stage are the following.
 The terminal velocity is given by
 \begin{equation} \label{w}
  \bm{w}=\alpha\bm{w}^{(2)}+\alpha^2\bm{w}^{(4)}+\mathrm{O}(\alpha^3)\;,
 \end{equation}
 with the leading order from (\ref{w2}) represented by:
 \begin{eqnarray} \label{W}
  \alpha\bm{w}^{(2)}\!\!&\!\!=\!\!&\!\!|1-\beta|\mathcal{J}\left[\tau\bm{g}+\inttx\bm{u}(\xt)\mathcal{J}^{-1}q^{(2)}(\xt)\right]\nonumber\\
  &\!\!=\!\!&\!\!\bm{\mathcal{W}}+(1-\beta)\!\int\!\bm{u}f(\bm{u},\tau,\bm{g},\mathcal{D})\;.
 \end{eqnarray}   
 Here we exploited the
 relation $1-\beta=\mathcal{J}\alpha$ and the fact that --- due to (\ref{ale}a) --- the field $\bm{r}^{(1)}/(\mathcal{J}\mathcal{K}^{-1})$ is independent
 of both $\mathcal{J}$ and $\mathcal{K}$ (i.e., of
 both $\beta$ and $\kappa$), so the same independence also holds for the field $q^{(2)}/\mathcal{J}$ because of (\ref{ale}b). Therefore, in the limit of
 quasi-neutrally-buoyant particles (and of $\kappa$ with the same order of smallness as $|1-\beta|$), the main contribution of the terminal velocity is
 represented by its bare value plus a \emph{same-order} deviation only dependent on the other finite quantities into play, and which can be computed
 numerically via (\ref{W})\&(\ref{ale}a--b). This leading order is independent of $\kappa$ and overall \emph{odd} in $1-\beta$: with all the other
 parameters fixed, particles slightly heavier than the fluid settle with a velocity opposite to the one of slightly lighter particles; at this stage, no
 immediate conclusion can be drawn on the sign of such a deviation. Note that expression (\ref{w}) does not exclude the possible presence of further terms
 linear in $|1-\beta|$ but with a ``prefactor'' proportional to a positive power of $\kappa$, because in our asymptotics these would be higher-order
 contributions. No immediate simplification can be performed on the term $\alpha^2\bm{w}^{(4)}$ from (\ref{w4}) for the time being.

 \section{Simplifications for flows endowed with vertical parity} \label{sec:pf}

 If a vertical-parity symmetry is imposed on the flow, further simplifications come along (at least for those situations where gravity is aligned with
 one side of the periodicity cell). Namely, if at a point $\bm{x}_*$ defined as the vertical reflection of the point $\bm{x}$ with respect to a
 reference horizontal plane ($\bm{x}_*\cdot\bm{g}=-\bm{x}\cdot\bm{g}$ and $\bm{x}_*\times\bm{g}=\bm{x}\times\bm{g}$), the vertical and
 horizontal components of the flow satisfy
 \begin{equation}
  \bm{u}(\bm{x}_*,t)\cdot\bm{g}=-\bm{u}(\xt)\cdot\bm{g}\quad\textrm{and}\quad\bm{u}(\bm{x}_*,t)\times\bm{g}=\bm{u}(\xt)\times\bm{g}\;,
 \end{equation}
 then it is possible to split all the relevant physical-space fields into their even and odd parts. For instance,
 $\bm{u}_{\mathrm{e/o}}(\xt)\equiv[\bm{u}(\xt)\pm\bm{u}(\bm{x}_*,t)]/2$, with a purely odd vertical component
 $\bm{u}\cdot\bm{g}=\bm{u}_{\mathrm{o}}\cdot\bm{g}$ and (a) purely even horizontal component(s) $\bm{u}\times\bm{g}=\bm{u}_{\mathrm{e}}\times\bm{g}$.
 The consequent equations derived from the sets (\ref{ale}) and (\ref{set}) are simpler to deal with, first of all from a numerical point of view as
 defined on a halved domain. Analytically, it can be shown that the function $f$ in (\ref{W}) is actually linear in $\bm{g}$, so that $\bm{w}^{(2)}$ is
 overall proportional to gravity; since the same can be stated also for $\bm{w}^{(4)}$ in (\ref{w}), such a conclusion holds for the whole
 terminal velocity at working order.
 
 Notice that this category also comprises cellular flows often adopted in analytical and numerical investigations to mimic Langmuir circulation on the
 ocean surface or lateral convective rolls in Rayleigh--B\'enard cells \cite{MC86,M87b,SG88,CMMV99,CMM00,MA08}.

 \section{Conclusions and perspectives} \label{sec:cp}
 
 We investigated the sedimentation process of quasi-neutrally buoyant inertial particles in zero-mean incompressible flows. Such particles are especially
 relevant in biophysical applications, where most of the aquatic micro-organisms \cite{LMAFMS12} have a mass density very similar to the one of water.
 General formulae have been found for their terminal velocity in generic space-and-time periodic (or steady) flows,
 with some additional information available for flows endowed with some degree of spatial symmetry such as negative parity in the vertical direction. 
 These expressions consist in space-time integrals of auxiliary quantities which satisfy partial differential equations of the
 advection--diffusion--reaction type, that can be solved at least numerically since our procedure allowed for a drastic reduction of the problem
 dimensionality from the full phase space to the classical physical space. Moreover, our
 expressions extend the range of validity of this approach to any value of the Stokes' time --- away from previous perturbative limits --- or at least to
 those situations where the basic dynamical system (\ref{basic}) makes sense and the (Basset, Oseen, Fax\'en, Saffman) corrections can be neglected.
 As a byproduct, our analysis also provides the physical-space particle probability density function once these differential equations are solved.
 
 Among the possible perspectives, first of all we mention the study of the particle effective --- or ``eddy'' --- diffusivity
 \cite{F95,AV95,BCVV95,ACMV00,PS05,MAMM12}. This can be performed by means of the multiple-scale method \cite{BO78,BLP78,PS07},
 and represents the following step in the investigation of higher-order effects in particle advection, including also the possibility of anomalous
 transport \cite{MA14,BMAM15}. When analyzing the possibility of a net displacement also in the horizontal direction, a clear connection with the problem
 of Stokes' drift arises \cite{S47,L86,SBMAMOP13}. Finally, we would like to attack the problem of particle dispersion following a point-source emission,
 an issue which has already been tackled for tracers \cite{C43,CMAM07} or slightly-inertial particles \cite{MAM11}, and that should be recast in the
 present framework of quasi-neutral buoyancy. 
 
 \paragraph{Acknowledgments} 
  We thank Andrea Mazzino, Paolo Muratore--Ginanneschi, Semyon Yakubovich and Luca Biferale for useful discussions and suggestions.
  This article is based upon work from COST Action MP1305, supported by COST (European Cooperation in Science and Technology).
  The authors were partially supported by CMUP (UID/MAT/00144/2013), which is funded by FCT (Portugal) with national (MEC) and European structural funds
  (FEDER), under the partnership agreement PT2020; and also by Project STRIDE - NORTE-01-0145-FEDER-000033, funded by ERDF — NORTE 2020.

 \appendix

 \section{Calculation details} \label{sec:ap}
 
 The first equation to attack is (\ref{01Ka}). Thanks to the full decoupling in the operator (\ref{l012a}), we can solve it through variable separation:
 \begin{eqnarray} \label{o0}
  &p^{(0)}(\xvt)=\sigma(\bm{y})q^{(0)}(\xt)\\
  &\displaystyle\Longrightarrow\frac{1}{q^{(0)}(\xt)}\mathcal{M}(\xt)q^{(0)}(\xt)=c=\frac{\tau^{-1}}{\sigma(\bm{y})}\mathcal{N}(\bm{y})\sigma(\bm{y})\;. \label{ab}
 \end{eqnarray}
 Looking at the right-hand equality in (\ref{ab}) integrated on the covelocity space, we get:
 \begin{eqnarray} \label{cov}
  &\displaystyle\!\int\!\ud\bm{y}\,c\sigma(\bm{y})=\tau^{-1}\!\int\!\ud\bm{y}\,\bm{\partial}_{\bm{y}}\cdot[\bm{y}\sigma(\bm{y})+\mathcal{K}\tau^{-1}\bm{\partial}_{\bm{y}}\sigma(\bm{y})]=0\Longrightarrow c=0\nonumber\\
  &\Longrightarrow \sigma(\bm{y})=(2\pi\mathcal{K}/\tau)^{-d/2}\mathrm{e}^{-y^2\tau/2\mathcal{K}}\textrm{ (chosen with unit normalization)\;.}
 \end{eqnarray}
 From the corresponding left-hand equality, we deduce an advection--diffusion equation in physical space:
 \begin{equation} \label{th0}
  \partial_tq^{(0)}(\xt)+\bm{u}(\xt)\cdot\bm{\partial}_{\bm{x}}q^{(0)}(\xt)-\mathcal{D}\partial^2_{\bm{x}}q^{(0)}(\xt)=0\;.
 \end{equation}
 For future use, we introduce the fully-symmetric polynomials (equivalent to multivariate $d$-dimensional Hermite polynomials,
 and with $+\mathcal{S}$ denoting the symmetrization process of any tensor on its free indices):
 \begin{eqnarray*}
  &\mathcal{C}_{ij}\equiv y_iy_j-\mathcal{K}\tau^{-1}\delta_{ij}\;,\qquad\mathcal{A}_{ijk}\equiv y_iy_jy_k-\mathcal{K}\tau^{-1}(y_i\delta_{jk}+\mathcal{S})\;,\\
  &\mathcal{B}_{ijkl}\equiv y_iy_jy_ky_l-\mathcal{K}\tau^{-1}(y_iy_j\delta_{kl}+\mathcal{S})+\mathcal{K}^2\tau^{-2}(\delta_{ij}\delta_{kl}+\mathcal{S})\;;\nonumber
 \end{eqnarray*}
 together with the Gaussian weight $\sigma(\bm{y})$, they enjoy the relations:
 \begin{eqnarray*}
  &\mathcal{N}(\bm{y})\sigma(\bm{y})=0\;,\quad\mathcal{N}(\bm{y})[y_i\sigma(\bm{y})]=-y_i\sigma(\bm{y})\;,\quad\mathcal{N}(\bm{y})[\mathcal{C}_{ij}\sigma(\bm{y})]=-2\mathcal{C}_{ij}\sigma(\bm{y})\;,\\
  &\mathcal{N}(\bm{y})[\mathcal{A}_{ijk}\sigma(\bm{y})]=-3\mathcal{A}_{ijk}\sigma(\bm{y})\;,\quad\mathcal{N}(\bm{y})[\mathcal{B}_{ijkl}\sigma(\bm{y})]=-4\mathcal{B}_{ijkl}\sigma(\bm{y})\;,\\
  &\displaystyle\!\int\!\ud\bm{y}\,y_i\sigma(\bm{y})=\!\int\!\ud\bm{y}\,\mathcal{C}_{ij}\sigma(\bm{y})=\!\int\!\ud\bm{y}\,\mathcal{A}_{ijk}\sigma(\bm{y})=\!\int\!\ud\bm{y}\,\mathcal{B}_{ijkl}\sigma(\bm{y})=0\;,\\
  &\displaystyle\textrm{along with }\!\int\!\ud\bm{y}\,\sigma(\bm{y})=1\;,\textrm{ and }\!\int\!\ud\bm{y}\,\bm{y}\otimes\bm{y}\sigma(\bm{y})=\mathcal{K}\tau^{-1}\mathsf{I}\;.
 \end{eqnarray*}
 By making use of lower-order results, we can now proceed to solve the system (\ref{01K}) recursively, starting from (\ref{01Kb}):
 \begin{eqnarray*}
  [\mathcal{M}(\xt)-\tau^{-1}\mathcal{N}(\bm{y})]p^{(1)}(\xyt)\!\!&\!\!=\!\!&\!\!-[\bm{y}\cdot\bm{\partial}_{\bm{x}}+\mathcal{J}\tau^{-1}\bm{z}(\xt)\cdot\bm{\partial}_{\bm{y}}]p^{(0)}(\xyt)\\
  &\!\!=\!\!&\!\!\sigma(\bm{y})y_i[-\partial_{x_i}+\mathcal{J}\mathcal{K}^{-1}z_i(\xt)]q^{(0)}(\xt)\;.
 \end{eqnarray*}
 The resolution passes through a process of Hermitianization (very closely related to the second-quantization algorithm \cite{MA08}). It consists in
 rewriting the unknown as the product between the Gaussian weight and an expansion in a power series in $\bm{y}$ up to the order in question, in this case
 the first, with space--time-dependent prefactors (notice that in (\ref{o0}) an expansion up to order 0, i.e.\ no expansion at all, appeared):
 \begin{equation} \label{o1}
  p^{(1)}(\xyt)=\sigma(\bm{y})[q^{(1)}(\xt)+y_ir^{(1)}_i(\xt)]
 \end{equation}
 \begin{equation}
  \Longrightarrow\left\{\begin{array}{rcl}
   [\partial_t+\bm{u}\cdot\bm{\nabla}-\mathcal{D}\nabla^2]q^{(1)}\!\!&\!\!=\!\!&\!\!0\\{}
   [\partial_t+\bm{u}\cdot\bm{\nabla}-\mathcal{D}\nabla^2+\tau^{-1}]r^{(1)}_i\!\!&\!\!=\!\!&\!\!-\nabla_iq^{(0)}+\mathcal{J}\mathcal{K}^{-1}z_iq^{(0)}\;.
  \end{array}\right. \label{th1}
 \end{equation}
 Resolution of (\ref{01Kc}) (for $\mathcal{I}=2$):
 \begin{eqnarray*}
  [\mathcal{M}(\xt)-\tau^{-1}\mathcal{N}(\bm{y})]p^{(2)}(\xyt)\!\!&\!\!=\!\!&\!\!-[\bm{y}\cdot\bm{\partial}_{\bm{x}}+\mathcal{J}\tau^{-1}\bm{z}(\xt)\cdot\bm{\partial}_{\bm{y}}]p^{(1)}(\xyt)\\
  &&\!\!+\mathcal{J}\bm{u}(\xt)\cdot\bm{\partial}_{\bm{x}}p^{(0)}(\xyt)
 \end{eqnarray*}
 \begin{equation} \label{o2}
  \Longrightarrow p^{(2)}(\xyt)=\sigma(\bm{y})[q^{(2)}(\xt)+y_ir^{(2)}_i(\xt)+\mathcal{C}_{ij}s^{(2)}_{ij}(\xt)]
 \end{equation}
 \begin{equation}
  \Longrightarrow\left\{\begin{array}{rcl}
   [\partial_t+\bm{u}\cdot\bm{\nabla}-\mathcal{D}\nabla^2]q^{(2)}\!\!&\!\!=\!\!&\!\!\mathcal{J}\bm{u}\cdot\bm{\nabla}q^{(0)}-\mathcal{K}\tau^{-1}\nabla_ir^{(1)}_i\\{}
   [\partial_t+\bm{u}\cdot\bm{\nabla}-\mathcal{D}\nabla^2+\tau^{-1}]r^{(2)}_i\!\!&\!\!=\!\!&\!\!-\nabla_iq^{(1)}+\mathcal{J}\mathcal{K}^{-1}z_iq^{(1)}\\{}
   [\partial_t+\bm{u}\cdot\bm{\nabla}-\mathcal{D}\nabla^2+2\tau^{-1}]s^{(2)}_{ij}\!\!&\!\!=\!\!&\!\!-\nabla_ir^{(1)}_j+\mathcal{J}\mathcal{K}^{-1}z_ir^{(1)}_j\;.
  \end{array}\right. \label{th2}
 \end{equation}
 Resolution of (\ref{01Kc}) (for $\mathcal{I}=3$):
 \begin{eqnarray*}
  [\mathcal{M}(\xt)-\tau^{-1}\mathcal{N}(\bm{y})]p^{(3)}(\xyt)\!\!&\!\!=\!\!&\!\!-[\bm{y}\cdot\bm{\partial}_{\bm{x}}+\mathcal{J}\tau^{-1}\bm{z}(\xt)\cdot\bm{\partial}_{\bm{y}}]p^{(2)}(\xyt)\\
  &&\!\!+\mathcal{J}\bm{u}(\xt)\cdot\bm{\partial}_{\bm{x}}p^{(1)}(\xyt)
 \end{eqnarray*}
 \begin{equation} \label{o3}
  \Longrightarrow p^{(3)}(\xyt)=\sigma(\bm{y})[q^{(3)}(\xt)+y_ir^{(3)}_i(\xt)+\mathcal{C}_{ij}s^{(3)}_{ij}(\xt)+\mathcal{A}_{ijk}a^{(3)}_{ijk}(\xt)]
 \end{equation}
 \begin{equation}
  \Longrightarrow\left\{\begin{array}{rcl}
   [\partial_t+\bm{u}\cdot\bm{\nabla}-\mathcal{D}\nabla^2]q^{(3)}\!\!&\!\!=\!\!&\!\!\mathcal{J}\bm{u}\cdot\bm{\nabla}q^{(1)}-\mathcal{K}\tau^{-1}\nabla_ir^{(2)}_i\\{}
   [\partial_t+\bm{u}\cdot\bm{\nabla}-\mathcal{D}\nabla^2+\tau^{-1}]r^{(3)}_i\!\!&\!\!=\!\!&\!\!-\nabla_iq^{(2)}+\mathcal{J}\mathcal{K}^{-1}z_iq^{(2)}+\mathcal{J}\bm{u}\cdot\bm{\nabla}r^{(1)}_i\\
   &&-\mathcal{K}\tau^{-1}\nabla_j(s^{(2)}_{ji}+s^{(2)}_{ij})\\{}
   [\partial_t+\bm{u}\cdot\bm{\nabla}-\mathcal{D}\nabla^2+2\tau^{-1}]s^{(3)}_{ij}\!\!&\!\!=\!\!&\!\!\ldots\\{}
   [\partial_t+\bm{u}\cdot\bm{\nabla}-\mathcal{D}\nabla^2+3\tau^{-1}]a^{(3)}_{ijk}\!\!&\!\!=\!\!&\!\!\ldots\;.
  \end{array}\right. \label{th3}
 \end{equation}
 Resolution of (\ref{01Kc}) (for $\mathcal{I}=4$):
 \begin{eqnarray*}
  [\mathcal{M}(\xt)-\tau^{-1}\mathcal{N}(\bm{y})]p^{(4)}(\xyt)\!\!&\!\!=\!\!&\!\!-[\bm{y}\cdot\bm{\partial}_{\bm{x}}+\mathcal{J}\tau^{-1}\bm{z}(\xt)\cdot\bm{\partial}_{\bm{y}}]p^{(3)}(\xyt)\\
  &&+\mathcal{J}\bm{u}(\xt)\cdot\bm{\partial}_{\bm{x}}p^{(2)}(\xyt)
 \end{eqnarray*}
 \begin{eqnarray} \label{o4}
  \Longrightarrow p^{(4)}(\xyt)=\sigma(\bm{y})[q^{(4)}(\xt)+y_ir^{(4)}_i(\xt)+\mathcal{C}_{ij}s^{(4)}_{ij}(\xt)+\mathcal{A}_{ijl}a^{(4)}_{ijl}(\xt)\nonumber\\
  +\mathcal{B}_{ijkl}b^{(4)}_{ijkl}(\xt)]
 \end{eqnarray}
 \begin{equation}
  \Longrightarrow\left\{\begin{array}{rcl}
   [\partial_t+\bm{u}\cdot\bm{\nabla}-\mathcal{D}\nabla^2]q^{(4)}\!\!&\!\!=\!\!&\!\!\mathcal{J}\bm{u}\cdot\bm{\nabla}q^{(2)}-\mathcal{K}\tau^{-1}\nabla_ir^{(3)}_i\\{}
   [\partial_t+\bm{u}\cdot\bm{\nabla}-\mathcal{D}\nabla^2+\tau^{-1}]r^{(4)}_i\!\!&\!\!=\!\!&\!\!\ldots\\{}
   [\partial_t+\bm{u}\cdot\bm{\nabla}-\mathcal{D}\nabla^2+2\tau^{-1}]s^{(4)}_{ij}\!\!&\!\!=\!\!&\!\!\ldots\\{}
   [\partial_t+\bm{u}\cdot\bm{\nabla}-\mathcal{D}\nabla^2+3\tau^{-1}]a^{(4)}_{ijk}\!\!&\!\!=\!\!&\!\!\ldots\\{}
   [\partial_t+\bm{u}\cdot\bm{\nabla}-\mathcal{D}\nabla^2+4\tau^{-1}]b^{(4)}_{ijkl}\!\!&\!\!=\!\!&\!\!\ldots\;.
  \end{array}\right. \label{th4}
 \end{equation}
 Note that, for our purpose, in (\ref{th3}) we only need to investigate $q^{(3)}$ and $\bm{r}^{(3)}$, and in (\ref{th4}) only $q^{(4)}$.
 It is also worth underlining that $q(\xt)=\sum_{\mathcal{I}=0}^{\infty}q^{(\mathcal{I})}(\xt)$, but the equations for the $q^{(\mathcal{I})}$'s
 necessarily imply the parallel resolution of the ones for the $\bm{r}^{(\bullet)}$'s and $\mathsf{s}^{(\bullet)}$'s to form a closed system and thus to
 compute the terminal velocity. 
 
 The overall normalization of the phase-space density $p$ corresponds to an integration on the whole covelocity space (either in the original coordinate
 $\bm{v}$ or in the rescaled one $\bm{y}$, which is indifferent because of the appearance of a Jacobian) and on the spatial periodicity cell, for any time:
 \begin{eqnarray} \label{norm}
  &\displaystyle\!\int_{\mathbb{P}}\!\ud\bm{x}\!\int_{\mathbb{R}^d}\!\ud\bm{v}\,p(\xvt)=1=\!\int_{\mathbb{P}}\!\ud\bm{x}\,q(\xt)\\
  &\displaystyle\Longrightarrow\!\int_{\mathbb{P}}\!\ud\bm{x}\!\int_{\mathbb{R}^d}\!\ud\bm{y}\,p^{(\mathcal{I})}(\xyt)=\delta_{\mathcal{I}0}=\!\int_{\mathbb{P}}\!\ud\bm{x}\,q^{(\mathcal{I})}(\xt)\;.\nonumber
 \end{eqnarray}
 For what concerns the initial conditions of $p$, they are more difficult to implement, nevertheless it is possible to impose them on $q^{(0)}$ and
 $q^{(1)}$. Indeed, these two scalar fields satisfy the unforced advection--diffusion equations (\ref{th0})\&(\ref{th1}a), whose unique periodic
 solutions (the one we are interested in) are the constants. The two exact values of the constants --- the inverse of the physical volume and zero,
 respectively --- are dictated by the spatial normalization (\ref{norm}) and by the covelocity one (\ref{cov}):
 \begin{equation} \label{th01}
  q^{(0)}(\bm{x},0)=\ell^{-d}=q^{(0)}(\xt)\;,\qquad q^{(1)}(\bm{x},0)=0=q^{(1)}(\xt)\;.
 \end{equation}
 Note that a transport property such as $\bm{w}$ cannot depend on the initial conditions, which are actually forgotten due to the diffusive term in the
 operator $\mathcal{M}$. In other frameworks where this independence is a priori not met, they must be taken as uniform or otherwise averaged upon.
 This point is strictly related to the fact that we neglect any possible transient decay in the phase-space density, and we only focus on its long-term
 behavior which influences the terminal velocity. This steady or periodic behavior of $p(\xvt)$ is due to the steady/periodic character of the fluid flow
 $\bm{u}(\xt)$, which is the only non-constant driving agent in the evolution equation (\ref{gfpkfk}).

 Keeping into account the expansions of the terms making up $p$ starting from (\ref{o0}), definition (\ref{dabliu}) translates into:
 \begin{equation} \label{w0}
  \bm{w}^{(0)}\!=\!\int_0^{\mathcal{T}}\!\frac{\ud t}{\mathcal{T}}\!\int_{\mathbb{P}}\!\ud\bm{x}\!\int_{\mathbb{R}^d}\!\ud\bm{y}\,\bm{u}(\xt)p^{(0)}(\xyt)=\inttx\bm{u}(\xt)q^{(0)}(\xt)=\bm{0}\;,
 \end{equation}
 \begin{eqnarray} \label{w1}
  \bm{w}^{(1)}\!\!&\!\!=\!\!&\!\!\displaystyle\!\int_0^{\mathcal{T}}\!\frac{\ud t}{\mathcal{T}}\!\int_{\mathbb{P}}\!\ud\bm{x}\!\int_{\mathbb{R}^d}\!\ud\bm{y}\,[\bm{u}(\xt)p^{(1)}(\xyt)+\bm{y}p^{(0)}(\xyt)]\nonumber\\
  &\!\!=\!\!&\!\!\displaystyle\inttx\bm{u}(\xt)q^{(1)}(\xt)=\bm{0}\;,
 \end{eqnarray}
 \begin{eqnarray} \label{wk}
  \bm{w}^{(\mathcal{I})}\!\!&\!\!=\!\!&\!\!\displaystyle\!\int_0^{\mathcal{T}}\!\frac{\ud t}{\mathcal{T}}\!\int_{\mathbb{P}}\!\ud\bm{x}\!\int_{\mathbb{R}^d}\!\ud\bm{y}\,\{\bm{u}(\xt)[p^{(\mathcal{I})}-\mathcal{\mathcal{J}}p^{(\mathcal{I}-2)}](\xyt)+\bm{y}p^{(\mathcal{I})}(\xyt)\}\nonumber\\
  &\!\!=\!\!&\!\!\displaystyle\inttx\{\bm{u}(\xt)[q^{(\mathcal{I})}-\mathcal{\mathcal{J}}q^{(\mathcal{I}-2)}](\xt)+\mathcal{K}\tau^{-1}\bm{r}^{(\mathcal{I}-1)}(\xt)\}
 \end{eqnarray}
 (for $\mathcal{I}\ge2$). The vanishing of expressions (\ref{w0}) and (\ref{w1}) is due to (\ref{th01}), in the former case coupled with (\ref{u}).
 Because of (\ref{th3}a) and (\ref{th2}b) (i.e.\ $q^{(3)}(\xt)=0$), one sees that also $\bm{w}^{(3)}=\bm{0}$, and similarly for all odd
 $\mathcal{I}$'s in (\ref{wk}) by induction.

 The relevant equations from the systems (\ref{th1})--(\ref{th4}) have already been reported in (\ref{ale}) and (\ref{set}).
 It is particularly useful to write down the temporal evolution of the following spatial integrals, arising from (\ref{th1}b) and (\ref{th3}b) respectively:
 \begin{equation} \label{r1}
  (\partial_t+\tau^{-1})\!\int_{\mathbb{P}}\!\ud\bm{x}\,r^{(1)}_i(\xt)=\mathcal{J}\mathcal{K}^{-1}\tau g_i\;,
 \end{equation}
 \begin{equation} \label{r3}
  (\partial_t+\tau^{-1})\!\int_{\mathbb{P}}\!\ud\bm{x}\,r^{(3)}_i(\xt)=\mathcal{J}\mathcal{K}^{-1}\!\int_{\mathbb{P}}\!\ud\bm{x}\,u_i(\xt)q^{(2)}(\xt)\;.
 \end{equation}
 A temporal integration of (\ref{r1}) allows us to recast (\ref{wk}) for $\mathcal{I}=2$ into the form (\ref{w2});
 a similar manipulation of (\ref{r3}) for $\mathcal{I}=4$ leads to (\ref{w4}).
 
 It is easy to show that parallel flows, i.e.\ fluid motions in which the velocity points always and everywhere in the same direction (say $x_1$), do not
 affect sedimentation if they are steady/periodic and incompressible --- implying that $\bm{u}$ does not depend on $x_1$ itself --- at least at working
 order. Indeed, for such a class of flows all the advective terms of the type $\bm{u}(\xt)\cdot\bm{\nabla}p(\xvt)$ vanish (also when acting on
 other statistical quantities based on $p$), because no long-term dependence on the spatial coordinate $x_1$ aligned with $\bm{u}$ can arise --- except for
 possible transient behaviors that can be neglected for our scope. As a consequence, one can easily prove that all the following quantities derived from
 (\ref{ale}) and (\ref{set}) vanish:
 \begin{equation*}
  \nabla\cdot\bm{r}^{(1)}(\xt)=q^{(2)}(\xt)=\nabla\nabla:\mathsf{s}^{(2)}(\xt)=\nabla\cdot\bm{r}^{(3)}(\xt)=q^{(4)}(\xt)=0\;.
 \end{equation*}
 Accordingly, $\bm{w}=\bm{\mathcal{W}}+O(\alpha^3)$, i.e.\ $\bm{\mathcal{Z}}=\bm{0}$ at working order.
 
 \subsection{Proof of the expression for the terminal-velocity correction} \label{sec:pr}
 
 Let us firstly prove the rewriting (\ref{dabliu}) of the full terminal velocity, by exploiting the definition of the phase-space density as an average
 of Dirac delta's on every random factor:
 \begin{equation} \label{pdelta}
  p(\xvt)\equiv\langle\delta(\bm{x}-\bm{\mathcal{X}}(t))\delta(\bm{v}-\bm{\mathcal{V}}(t))\rangle_{\bm{\mu},\bm{\nu}}\;.
 \end{equation}
 Let us remind that both $\bm{\mu}(t)$ and $\bm{\nu}(t)$ are white noises, meaning that the values assumed at a certain time instant are completely
 uncorrelated from the ones assumed at a time instant immediately following. Moreover, by invoking causality, one infers that the instantaneous values
 of the noises at time $t$ can influence the particle dynamics only at future times, but not computed at $t$ itself. This means that
 \begin{equation} \label{munu}
  \langle\bm{\mu}(t)\delta(\bm{x}-\bm{\mathcal{X}}(t))\delta(\bm{v}-\bm{\mathcal{V}}(t))\rangle_{\bm{\mu},\bm{\nu}}=\bm{0}=\langle\bm{\nu}(t)\delta(\bm{x}-\bm{\mathcal{X}}(t))\delta(\bm{v}-\bm{\mathcal{V}}(t))\rangle_{\bm{\mu},\bm{\nu}}\;,
 \end{equation}
 since the averages split thanks to the uncorrelation, and both noises have zero mean.\\
 For the sake of notational simplicity, let
 \begin{equation*}
  \itxv\bullet\equiv\inttxv\bullet\;.
 \end{equation*}
 As a consequence, using (\ref{munu}) and (\ref{pdelta}),
 \begin{eqnarray*}
  &&\!\!\langle\bm{\mathcal{V}}(t)+\beta\bm{u}(\bm{\mathcal{X}}(t),t)+\sqrt{2\mathcal{D}}\bm{\mu}(t)\rangle_{p}\\
  &\!\!=\!\!&\!\!\itxv\langle[\bm{\mathcal{V}}(t)+\beta\bm{u}(\bm{\mathcal{X}}(t),t)+\sqrt{2\mathcal{D}}\bm{\mu}(t)]\delta(\bm{x}-\bm{\mathcal{X}}(t))\delta(\bm{v}-\bm{\mathcal{V}}(t))\rangle_{\bm{\mu},\bm{\nu}}\\
  &\!\!=\!\!&\!\!\itxv\langle[\bm{v}+\beta\bm{u}(\xt)]\delta(\bm{x}-\bm{\mathcal{X}}(t))\delta(\bm{v}-\bm{\mathcal{V}}(t))\rangle_{\bm{\mu},\bm{\nu}}+\bm{0}\\
  &\!\!=\!\!&\!\!\itxv[\bm{v}+\beta\bm{u}(\xt)]p(\xvt)\;,
 \end{eqnarray*}
 thanks to the property of the delta which allows for the substitution $(\bm{\mathcal{X}}(t),\bm{\mathcal{V}}(t))\mapsto(\bm{x},\bm{v})$,
 and to the fact that these latter coordinates are independent of the noises.\\
 Let us now compute the deviation (\ref{delta}) of the terminal velocity (\ref{dabliu}) from its bare value (\ref{ast}):
 \begin{eqnarray*}
  \bm{\mathcal{Z}}\!\!&\!\!=\!\!&\!\!\itxv[\bm{v}+\beta\bm{u}(\xt)]p(\xvt)-(1-\beta)\tau\bm{g}\\
  &\!\!=\!\!&\!\!\itxv[\bm{v}+\beta\bm{u}(\xt)-(1-\beta)\tau\bm{g}]\langle\delta(\bm{x}-\bm{\mathcal{X}}(t))\delta(\bm{v}-\bm{\mathcal{V}}(t))\rangle_{\bm{\mu},\bm{\nu}}\\
  &\!\!=\!\!&\!\!\langle\itxv[\bm{\mathcal{V}}(t)+\beta\bm{u}(\bm{\mathcal{X}}(t),t)-(1-\beta)\tau\bm{g}]\delta(\bm{x}-\bm{\mathcal{X}}(t))\delta(\bm{v}-\bm{\mathcal{V}}(t))\rangle_{\bm{\mu},\bm{\nu}}\;,
 \end{eqnarray*}
 having used (\ref{norm}). Exploiting the second equation of (\ref{basic}), this rewrites as:
 \begin{eqnarray*}
  \bm{\mathcal{Z}}\!\!&\!\!=\!\!&\!\!\langle\itxv[\bm{u}(\bm{\mathcal{X}}(t),t)+\sqrt{2\kappa}\bm{\nu}(t)-\tau\dot{\bm{\mathcal{V}}}(t)]\delta(\bm{x}-\bm{\mathcal{X}}(t))\delta(\bm{v}-\bm{\mathcal{V}}(t))\rangle_{\bm{\mu},\bm{\nu}}\\
  &\!\!=\!\!&\!\!\itxv\bm{u}(\xt)p(\xvt)+\bm{0}-\tau\langle\itxv\dot{\bm{\mathcal{V}}}(t)\delta(\bm{x}-\bm{\mathcal{X}}(t))\delta(\bm{v}-\bm{\mathcal{V}}(t))\rangle_{\bm{\mu},\bm{\nu}}\;,
 \end{eqnarray*}
 after making use of (\ref{munu}). Keeping (\ref{pP}) in mind, the demonstration is complete if we prove that the addend involving $\dot{\bm{\mathcal{V}}}$
 does not give any contribution. This is achieved through integrations by parts (with vanishing of the integrals of derivatives, because of periodicity and
 rapid decay at infinity) and chain-rule derivation, and the exploitation of the material and functional derivatives --- $\ud$ and $\mathrm{D}$
 respectively --- and of the translational invariance of the delta's:
 \begin{eqnarray*}
  &&\!\!\langle\itxv\dot{\bm{\mathcal{V}}}(t)\delta(\bm{x}-\bm{\mathcal{X}}(t))\delta(\bm{v}-\bm{\mathcal{V}}(t))\rangle_{\bm{\mu},\bm{\nu}}\\
  &\!\!=\!\!&\!\!\left\langle\itxv\left\{\frac{\ud}{\ud t}[\bm{\mathcal{V}}(t)\delta(\bm{x}-\bm{\mathcal{X}}(t))\delta(\bm{v}-\bm{\mathcal{V}}(t))]-\bm{\mathcal{V}}(t)\frac{\ud}{\ud t}[\delta(\bm{x}-\bm{\mathcal{X}}(t))\delta(\bm{v}-\bm{\mathcal{V}}(t))]\right\}\right\rangle_{\bm{\mu},\bm{\nu}}\\
  &\!\!=\!\!&\!\!\itxv\frac{\partial}{\partial t}\langle\bm{\mathcal{V}}(t)\delta(\bm{x}-\bm{\mathcal{X}}(t))\delta(\bm{v}-\bm{\mathcal{V}}(t))\rangle_{\bm{\mu},\bm{\nu}}-\left\langle\itxv\bm{\mathcal{V}}(t)\bigg[\delta(\bm{v}-\bm{\mathcal{V}}(t))\dot{\bm{\mathcal{X}}}(t)\cdot\right.\\
  &&\left.\left.\cdot\frac{\mathrm{D}}{\mathrm{D}\bm{\mathcal{X}}(t)}\delta(\bm{x}-\bm{\mathcal{X}}(t))+\delta(\bm{x}-\bm{\mathcal{X}}(t))\dot{\bm{\mathcal{V}}}(t)\cdot\frac{\mathrm{D}}{\mathrm{D}\bm{\mathcal{V}}(t)}\delta(\bm{v}-\bm{\mathcal{V}}(t))\right]\right\rangle_{\bm{\mu},\bm{\nu}}\\
  &\!\!=\!\!&\!\!0+\left\langle\itxv\bm{\mathcal{V}}(t)\left[\delta(\bm{v}-\bm{\mathcal{V}}(t))\dot{\bm{\mathcal{X}}}(t)\cdot\frac{\partial}{\partial\bm{x}}\delta(\bm{x}-\bm{\mathcal{X}}(t))\right.\right.\\
  &&\left.\left.+\delta(\bm{x}-\bm{\mathcal{X}}(t))\dot{\bm{\mathcal{V}}}(t)\cdot\frac{\partial}{\partial\bm{v}}\delta(\bm{v}-\bm{\mathcal{V}}(t))\right]\right\rangle_{\bm{\mu},\bm{\nu}}\\
  &\!\!=\!\!&\!\!\itxv\left[\frac{\partial}{\partial\bm{x}}\cdot\langle\dot{\bm{\mathcal{X}}}(t)\delta(\bm{x}-\bm{\mathcal{X}}(t))\delta(\bm{v}-\bm{\mathcal{V}}(t))\bm{\mathcal{V}}(t)\rangle_{\bm{\mu},\bm{\nu}}\right.\\
  &&\left.+\frac{\partial}{\partial\bm{v}}\cdot\langle\dot{\bm{\mathcal{V}}}(t)\delta(\bm{x}-\bm{\mathcal{X}}(t))\delta(\bm{v}-\bm{\mathcal{V}}(t))\bm{\mathcal{V}}(t)\rangle_{\bm{\mu},\bm{\nu}}\right]\\
  &\!\!=\!\!&\!\!0+0\qquad\textrm{Q.E.D.}
 \end{eqnarray*}

 \paragraph{References}


\begin{thebibliography}{99}
  \bibitem{R90} M.W. Reeks, On a kinetic equation for the transport of particles in turbulent flows, Phys.\ Fluids A 3(3) (1991) 446--456.
  \bibitem{R92} M.W. Reeks, On the continuum equations for dispersed particles in nonuniform flows, Phys.\ Fluids A 4(6) (1992) 1290--1303.
  \bibitem{BFF01} E. Balkovsky, G. Falkovich, A. Fouxon, Intermittent distribution of inertial particles in turbulent flows, Phys.\ Rev.\ Lett.\ 86 (2001) 2790--2793.
  \bibitem{B03} J. Bec, Fractal clustering of inertial particles in random flows, Phys.\ Fluids 15 (2003) L81--L84.
  \bibitem{WM03} M. Wilkinson, B. Mehlig, Path coalescence transition and its applications, Phys.\ Rev.\ E 68 (2003) 040101(R):1--4.
  \bibitem{FP04} G. Falkovich, A. Pumir, Intermittent distribution of heavy particles in a turbulent flow, Phys.\ Fluids 16 (2004) L47--50.
  \bibitem{ML04} I.M. Mazzitelli, D. Lohse, Lagrangian statistics for fluid particles and bubbles in turbulence, New J.\ Phys.\ 6 (2004) 1--28.
  \bibitem{CBBBCLMT06} M. Cencini, J. Bec, L. Biferale, G. Boffetta, A. Celani, A.S. Lanotte, S. Musacchio, F. Toschi, Dynamics and statistics of heavy particles in turbulent flows, J.\ Turb.\ 7 (2006) 36:1--36.
  \bibitem{VCVLMPT08} R. Volk, E. Calzavarini, G. Verhille, D. Lohse, N. Mordant, J.-F. Pinton, F. Toschi, Acceleration of heavy and light particles in turbulence: Comparison between experiments and direct numerical simulations, Physica D 237 (2008) 2084--2089.
  \bibitem{KPSTT00} G. K\'arolyi, \'A. P\'entek, I. Scheuring, T. T\'el, Z. Toroczkai, Chaotic flow: the physics of species coexistence, Proc.\ Natl Acad.\ Sci.\ 97 (2000) 13661--13665.
  \bibitem{HDS08} C. Habchi, N. Dumont, O. Simonin, Multidimensional simulation of cavitating flows in diesel injectors by a homogeneous mixture modeling approach, Atomiz.\ Spr.\ 18(2) (2008) 129--162.
  \bibitem{MM02} S. Matarrese, R. Mohayee, The growth of structure in the intergalactic medium, Mon.\ Not.\ R.\ Astron.\ Soc.\ 329 (2002) 37--60.
  \bibitem{FFS02} G. Falkovich, A. Fouxon, M. Stepanov, Acceleration of rain initiation by cloud turbulence, Nature 419 (2002) 151--154.
  \bibitem{MY75} A.S. Monin, A.M. Yaglom, Statistical Fluid Mechanics, MIT Press, Cambridge, 1975.
  \bibitem{AM91} M. Avellaneda, A. Majda, An integral representation and bounds on the effective diffusivity in passive advection and turbulent flows, Commun.\ Math.\ Phys.\ 138 (1991) 339--391.
  \bibitem{HM94} D.J. Horntrop, A. Majda, Subtle statistical behavior in simple models for random advection--diffusion, J.\ Math.\ Sci.\ Univ.\ Tokyo 1 (1994) 23--70.
  \bibitem{MV97} A. Mazzino, M. Vergassola, Interference between turbulent and molecular diffusion, Europhys.\ Lett.\ 37(8) (1997) 535--540.
  \bibitem{MMV05} A. Mazzino, S. Musacchio, A. Vulpiani, Multiple-scale analysis and renormalization for preasymptotic scalar transport, Phys.\ Rev.\ E 71 (2005) 011113: 1--11.
  \bibitem{CMMV06} M. Cencini, A. Mazzino, S. Musacchio, A. Vulpiani, Large-scale effects on meso-scale modeling for scalar transport, Physica D 220 (2006) 146--156.
  \bibitem{FN10} R. Ferrari, M. Nikurashin, Suppression of Eddy Diffusivity across Jets in the Southern Ocean, J.\ Phys.\ Oceanogr.\ 40 (2010) 1501--1519.
  \bibitem{MAMG16} M. Martins Afonso, A. Mazzino, S. Gama, Combined role of molecular diffusion, mean streaming and helicity in the eddy diffusivity of short-correlated random flows, J.\ Stat.\ Mech. 2016(10) (2016) 103205:1--17.
  \bibitem{MR83} M.R. Maxey, J.J. Riley, Equation of motion for a small rigid sphere in a nonuniform flow, Phys. Fluids 26(4) (1983) 883--889.
  \bibitem{G83} R. Gatignol, The Fax\'en formulae for a rigid particle in an unsteady non-uniform Stokes flow, J.\ M\'ec.\ Th\'eor.\ Appl. 1 (1983) 143--160.
  \bibitem{R88} M.W. Reeks, The relationship between Brownian motion and the random motion of small particles in a turbulent flow, Phys.\ Fluids 31 (1988) 1314--1316.
  \bibitem{C43} S. Chandrasekhar, Stochastic problems in physics and astronomy, Rev.\ Mod.\ Phys.\ 15 (1943) 1--89.
  \bibitem{G85} C.W. Gardiner, Handbook of Stochastic Methods: for Physics, Chemistry and the Natural Sciences, Springer, Berlin, 1985.
  \bibitem{R89} H. Risken, The Fokker--Planck Equation: Methods of Solutions and Applications, Springer, Berlin, 1989.
  \bibitem{V07} N.G. Van Kampen, Stochastic Processes in Physics and Chemistry, Elsevier,  Amsterdam, 2007.
  \bibitem{MC86} M.R. Maxey, S. Corrsin, Gravitational settling of aerosol particles in randomly oriented cellular flow fields, J.\ Atmos.\ Sci.\ 43(11) (1986) 1112--1134.
  \bibitem{M87a} M.R. Maxey, The gravitational settling of aerosol particles in homogeneous turbulence and random flow fields, J.\ Fluid Mech.\ 174 (1987) 441--465.
  \bibitem{M87b} M.R. Maxey, The motion of small spherical particles in a cellular flow field, Phys.\ Fluids 30 (1987) 1915--1928.
  \bibitem{WM93} L.P. Wang, M.R. Maxey, Settling velocity and concentration distribution of heavy particles in homogeneous isotropic turbulence, J.\ Fluid Mech.\ 256 (1993) 27--68.
  \bibitem{FK02} P.D. Friedman, J. Katz, Mean rise rate of droplets in isotropic turbulence, Phys. Fluids 14 (2002) 3059--3073.
  \bibitem{RMP04} J. Ruiz, D. Mac\'{\i}as, P. Peters, Turbulence increases the average settling velocity of phytoplankton cell, Proc.\ Natl Acad.\ Sci.\ 101 (2004) 17720--17724.
  \bibitem{MA08} M. Martins Afonso, The terminal velocity of sedimenting particles in a flowing fluid, J.\ Phys.\ A 41(38) (2008) 385501:1--15.
  \bibitem{BCPP00} A. Babiano, J.H.E. Cartwright, O. Piro, A. Provenzale, Dynamics of a Small Neutrally Buoyant Sphere in a Fluid and Targeting in Hamiltonian Systems, Phys.\ Rev.\ Lett.\ 84 (2000) 5764--5767.
  \bibitem{MFS07} C. Marchioli, M. Fantoni, A. Soldati, Influence of added mass on anomalous high rise velocity of light particles in cellular flow field: a note on the paper by Maxey (1987), Phys.\ Fluids 19 (2007) 098101:1--4.
  \bibitem{SH08} T. Sapsis, G. Haller, Instabilities in the dynamics of neutrally buoyant particles, Phys.\ Fluids 20 (2008) 017102:1--7.
  \bibitem{CC99} P. Castiglione, A. Crisanti, Dispersion of passive tracers in a velocity field with non-$\delta$-correlated noise, Phys.\ Rev.\ E 59(4) (1999) 3926--3934.
  \bibitem{BMM17} S. Boi, A. Mazzino, P. Muratore-Ginanneschi, Eddy diffusivities of inertial particles in random Gaussian flows, Phys.\ Rev.\ Fluids 2 (2017) 014602:1--8.
  \bibitem{MAM11} M. Martins Afonso, A. Mazzino, Point-source inertial particle dispersion, Geophys.\ Astrophys.\ Fluid Dyn.\ 105(6) (2011) 553--565.
  \bibitem{MAMM12} M. Martins Afonso, A. Mazzino, P. Muratore-Ginanneschi, Eddy diffusivities for inertial particles under gravity, J.\ Fluid Mech.\ 694 (2012) 426--463.
  \bibitem{SG88} T.H. Solomon, J.P. Gollub, Chaotic particle transport in time-dependent Rayleigh--B\'enard convection, Phys.\ Rev.\ A 38 (1988) 6280--6286.
  \bibitem{CMMV99} P. Castiglione, A. Mazzino, P. Muratore-Ginanneschi, A. Vulpiani, On strong anomalous diffusion, Physica D 134 (1999) 75--93.
  \bibitem{CMM00} P. Castiglione, A. Mazzino, P. Muratore--Ginanneschi, Numerical study of strong anomalous diffusion. Physica A 280 (2000) 60--68.
  \bibitem{LMAFMS12} M. Link\`es, M. Martins Afonso, P. Fede, J. Morchain, P. Schmitz, Numerical study of substrate assimilation by a microorganism exposed to fluctuating concentration, Chem.\ Eng.\ Sci.\ 81 (2012) 8--19. 
  \bibitem{F95} U. Frisch, Turbulence, Cambridge University Press, Cambridge, 1995.
  \bibitem{AV95} M. Avellaneda, M. Vergassola, Stieltjes integral representation of effective diffusivities in time-dependent flows, Phys.\ Rev.\ E 52(3) (1995) 3249--3251.
  \bibitem{BCVV95} L. Biferale, A. Crisanti, M. Vergassola, A. Vulpiani, Eddy diffusivities in scalar transport, Phys.\ Fluids 7(11) (1995) 2725--2734.
  \bibitem{ACMV00} K.H. Andersen, P. Castiglione, A. Mazzino, A. Vulpiani, Simple stochastic models showing strong anomalous diffusion, Eur.\ Phys.\ J.\ B 18 (2000) 447--452.
  \bibitem{PS05} G.A. Pavliotis, A.M. Stuart, Periodic homogenization for inertial particles, {\it Physica} D 204 (2005) 161--187.
  \bibitem{BO78} C.M. Bender, S.A. Orszag, Advanced Mathematical Methods for Scientists and Engineers, McGraw-Hill, New York, 1978.
  \bibitem{BLP78} A. Bensoussan, J.-L. Lions, G. Papanicolaou, Asymptotic Analysis of Periodic Structures, North-Holland, Amsterdam, 1978.
  \bibitem{PS07} G.A. Pavliotis, A.M. Stuart, Multiscale Methods: Averaging and Homogenization, Springer, Berlin, 2007.
  \bibitem{MA14} M. Martins Afonso, Anomalous diffusion for inertial particles under gravity in parallel flows, Phys.\ Rev.\ E 89(6) (2014) 063021:1--8.
  \bibitem{BMAM15} S. Boi, M. Martins Afonso, A. Mazzino, Anomalous diffusion of inertial particles in random parallel flows: theory and numerics face to face, J.\ Stat.\ Mech.\ 2015(10) (2015) P10023:1--21.
  \bibitem{S47} G.G. Stokes, On the theory of oscillatory waves, Trans.\ Cambridge Philos.\ Soc.\ 8 (1847) 441--473.
  \bibitem{L86} M.S. Longuet-Higgins, Eulerian and Lagrangian aspects of surface waves, J.\ Fluid Mech.\ 173 (1986) 683--707.
  \bibitem{SBMAMOP13} F. Santamaria, G. Boffetta, M. Martins Afonso, A. Mazzino, M. Onorato, D. Pugliese, Stokes drift for inertial particles transported by water waves, Europhys.\ Lett.\ 102(1) (2013) 14003:1--5.
  \bibitem{CMAM07} A. Celani, M. Martins Afonso, A. Mazzino, Point-source scalar turbulence, J.\ Fluid Mech.\ 583 (2007) 189--198.
 \end{thebibliography}
\end{document}